 \documentclass[preprint,showpacs,showkeys,preprintnumbers,amsmath,amssymb]{revtex4}

\usepackage{graphicx}  
\usepackage{dcolumn}   
\usepackage{bm}	       

\newcommand{\D}{\partial}
\newcommand{\ve}[1]{{\bf\underline #1}}
\newcommand{\vve}[1]{{\bf\underline{\underline #1}}}

\newcommand{\Dt}[1]{\frac {\D #1} {\D t}}

\newcommand{\dt}[1]{\frac {d#1} {d t}}

\newcommand{\bracket}[1]{\left[#1\right]}

\newcommand{\parenth}[1]{\left(#1\right)}
\newcommand{\II}{I\!\!I}
\DeclareMathSymbol{\R}{\mathbin}{AMSb}{"52}

\begin{document}

	\title{Remarkable evolutionary laws of absolute and relative
	entropies with dynamical systems}

	\author{X. San Liang}
	\email{sanliang@courant.nyu.edu}
	\affiliation{China Institute for Advanced Study,
		Central University of Finance and Economics, 
		Beijing, China 100081,\\
		and Courant Institute of Mathematical Sciences \\
			New York, New York 10012}
 	\date{July 4, 2010}

\begin{abstract}

The evolution of entropy is derived with respect to dynamical systems. 
For a stochastic system, its relative entropy $D$ evolves in accordance 
with the second law of thermodynamics; its absolute entropy $H$
may also be so, provided that the stochastic perturbation is additive 
and the flow of 
the vector field is nondivergent. For a deterministic system, 
$dH/dt$ is equal to the mathematical expectation of the divergence 
of the flow (a result obtained before), and, remarkably, $dD/dt = 0$. 
That is to say, relative entropy is always conserved.
So, for a nonlinear system, though the trajectories of the 
state variables, say $\ve x$, may appear chaotic in the phase space, 
say $\Omega$, those of the density function $\rho(\ve x)$
in the new ``phase space'' $L^1(\Omega)$ are not; the corresponding
Lyapunov exponent is always zero. This result is expected to have
important implications for the ensemble predictions in many applied fields,
and may help to analyze chaotic data sets.

\end{abstract}




 \maketitle


In thermodynamics, it is well known that entropy production is closely
related to phase space contraction\cite{thermo_entropy}. 
In information theory, similar relation has also been established; 
particularly, in the context of a deterministic system, it has been
shown that the time evolution of absolute entropy, 
namely Shannon entropy, is precisely equal to the 
mathematical expectation of the divergence of the flow
(cf.~Eq.~(\ref{eq:entropy_gov}) below)\cite{LK05}. 
This elegant relation has since led to the establishment of 
a rigorous formalism of information flow (or information transfer as
referred to in the literature), a fundamental notion in general phsyics
which has broad applications in a variety of disciplines\cite{LK05}\cite{Liang}. 

However, it has also been well known that absolute entropy, 
denoted $H$ henceforth, need not be consistent with the 
second law of thermodynamics\cite{Cove}
which states that the entropy of an isolated system cannot decrease as 
time goes on. Although the connection between information entropy and 
thermodynamic entropy is still on debate\cite{debate}, 
it would be better to have the former put on a physical footing. 
In this case, naturally one would ask under what circumstances the 
consistency may be established. This forms one of the questions
we want to address in this study.

On the other hand, relative entropy (hereafter $D$) does comply with the
second law of thermodynamics\cite{Cove}.
This important property, among others, makes $D$ an ideal physical measure
in many contexts, as recognized by Kleeman (2002), and has let to a 
resurgence of interest in it during the past decade in different 
applications\cite{rela_entropy}.
Considering that $H$ has a concise evolutionary law, 
one naturally wonders how $D$ evolves. In \cite{Cove}, this
is discussed in the framework of a Markov chain, and obtained is an
inequality like the afore-mentioned second law. 
But somehow the result is too generic; 
in the context of a dynamical system, it could have a more
specific and, hopefully, more definite statement.
Indeed, as we will see soon, relative entropy is actually 
conserved with deterministic systems. This remarkable result,
together with others, are what we are about to derive in the following.


First consider an $n$-dimensional deterministic system with 
randomness limited within initial conditions: 
	\begin{eqnarray}	\label{eq:gov}
	\dt {\ve x} = \ve F(\ve x, t),
	\end{eqnarray}
where $\ve x = (x_1, x_2, ..., x_n)^T \in \R^n$ are the state variables.
Associated with $\ve x$ there is a joint probability density function,
$\rho = \rho(t; \ve x) = \rho(t; x_1, x_2, ..., x_n)$, and hence 
an absolute entropy 
	\begin{eqnarray}	\label{eq:H}
	H = - \int_{\R^n} \rho\log\rho\ d\ve  x,
	\end{eqnarray}
and a relative entropy 
	\begin{eqnarray}	\label{eq:D}
	D = \int_{\R^n} \rho \log \frac \rho q\ d\ve x
	  = - H - \int_{\R^n} \rho \log q\ d\ve x,
	\end{eqnarray}
with some reference density $q$ of $\ve x$.
We are interested in how $H$ and $D$ evolve with respect to (\ref{eq:gov}).
For this purpose, assume that $\rho$, $q$, and their derviatives are 
all compactly supported; further assume enough regularity for 
$\rho$, $q$, $D$, and $H$. 
The mathematics involved here is neglected for a broad readership;
those who feel interested may consult \cite{Cove} for a detailed discussion.
Note the choosing of the reference density $q$ is slightly different 
from what people are using these days\cite{rela_entropy}
in applications, particularly in predictability studies, who usually
choose it to be some constant distribution (initial distribution, for example).
We require that $q$ also evolve, and that it
follow the same evolution as $\rho$ does. Only in this way
can we have the neat result on $D$, as will be derived soon.
(Perhaps this is the reason why the following result was not seen
before, as the past studies have focused on the choice of a constant $q$.)

Corresponding to (\ref{eq:gov}) there is a Liouville equation
	\begin{eqnarray}	\label{eq:liouville}
	\Dt\rho + \nabla\cdot(\rho\ve F) = 0
	\end{eqnarray}
governing the evolution of the joint density $\rho$.
Multiplying (\ref{eq:liouville}) by $-(1+\log\rho)$ and integrating over
$\R^n$, Liang and Kleeman obtain that\cite{LK05} 
    \begin{center}
      \framebox[0.4\textwidth]{
      \begin{minipage}[c]{1\textwidth}
        \begin{eqnarray}
        \dt {H} =  E \parenth{\nabla\cdot\ve F}, \label{eq:entropy_gov}\cr
        \end{eqnarray}
      \end{minipage}
      }
    \end{center}
where the operator $E$ stands for mathematical expectation
(refer to \cite{LK05} for the derivation).
In arriving at this formula, originally it is assumed that extreme
events have a probability of zero, which corresponds to our above compact
support assumption. This makes sense in practice and has
been justified in \cite{LK05}, but even this assumption may be relaxed,
and the same formula follows\cite{Liang}.

For the relative entropy $D$, differentiation
of (\ref{eq:D}) with respect to $t$ gives
	\begin{eqnarray*}
	\dt D &=& - \dt H - \int \Dt\rho \log q\ d\ve x
			  - \int \frac\rho q \Dt q\ d\ve x	\cr
	      &\equiv& - \dt H + (I) + (\II).
	\end{eqnarray*}
The integrals are all understood to be over $\R^n$, and this simplification
will be used hereafter, unless otherwise indicated.
The two shorthands are:
	\begin{eqnarray*}
	(I) &=& \int_\Omega \bracket{\nabla\cdot(\rho\ve F) \log q}\ d\ve x
	    = \int_\Omega \nabla\cdot (\rho\ve F \log q) -
	      \int_\Omega \rho\ve F\cdot \nabla(\log q)		\cr
	    &=& - E \parenth{\ve F \cdot \nabla\log q},		\\
	(\II) &=& - E\parenth{\Dt {\log q}}.
	\end{eqnarray*}
So 
	\begin{eqnarray}	\label{eq:tmp1}
	\dt D &=& - \dt H - E \parenth{\Dt{\log q} + \ve F\cdot\nabla\log q} \cr
	      &=& - \dt H - E \bracket{\frac 1 q 
			      \parenth{\Dt q + \ve F\cdot\nabla q}}.
	\end{eqnarray}
Recall that $q$ is also a joint density of $\ve x$,
so its evolution must follow the same Liouville equation, i.e.,
	\begin{eqnarray*}
        \Dt q + \ve F\cdot\nabla q = - q \nabla\cdot \ve F.
	\end{eqnarray*}
The relative entropy evolution (\ref{eq:tmp1}) thus becomes
	\begin{eqnarray}	
	\dt D = - \dt H + E(\nabla\cdot\ve F).
	\end{eqnarray}
Substitution of (\ref{eq:entropy_gov}) for $\dt H$ gives
    \begin{center}
      \framebox[0.4\textwidth]{
      \begin{minipage}[c]{1\textwidth}
        \begin{eqnarray}
        \dt D =  0. 		\label{eq:re_entropy_gov}\cr
        \end{eqnarray}
      \end{minipage}
      }
    \end{center}
That is to say, relative entropy is conserved.



The above results are now generalized to systems with stochasticity included.
Let $\ve w = (w_1, w_2, ..., w_n)^T$ be an array of $n$ 
standard Wiener processes, and $\vve B$ a matrix which may 
have dependency on both $\ve x$ and $t$. The system we are to consider
has the form:
	\begin{eqnarray}	\label{eq:gov2}
	d\ve x = \ve F(\ve x, t)dt + \vve B(\ve x,t) d \ve w.
	\end{eqnarray}
Correspondingly the density evolves according to a
Fokker-Planck equation	
	\begin{eqnarray} 	\label{eq:fokker}
	\Dt\rho = - \nabla\cdot {(\rho \ve F)} + 
	\frac 1 2 \nabla\nabla : (\rho\vve G),
	\end{eqnarray}
%
%
where $\vve G = \vve B\ \vve B^T$ is a nonnegatively definite matrix. 
  The double dot product here is defined
  such that, for column vectors $\ve a$, $\ve b$, $\ve c$, and $\ve d$, 
	$$(\ve a \ve b) : (\ve c \ve d) = 
	  (\ve a \cdot \ve c) (\ve b \cdot \ve d).$$
  A dyad $\ve a \ve b$ in matrix notation is identified with $\ve a \ve b^T$.

Multiplication of (\ref{eq:fokker}) by $-(1+\log\rho)$, followed by an 
integration over the entire sample space $\R^n$, yields an evolution 
of the absolute entropy
	\begin{eqnarray}	\label{eq:fokker_H}
	\dt H = E\parenth{\nabla \cdot \ve F} - \frac 1 2
        \int (1+\log\rho) \nabla\nabla : (\rho\vve G)\ d\ve x.
	\end{eqnarray}
In arriving at the first term on the right hand, 
the previous result (i.e., (\ref{eq:entropy_gov})) 
with the Liouville equation has been applied.
For the second term, since $\int \nabla\nabla : (\rho\vve G) d\ve x = 0$
by the compact support assumption, it results in
     \begin{eqnarray*}
     && -\frac 1 2 \int \log\rho \nabla\nabla : (\rho\vve G)\ d\ve x
     =\frac 12 \int \nabla\cdot (\rho\vve G) \cdot \nabla\log\rho\ d\ve x\\
     &&\qquad = - \frac 12 \int \rho\vve G : \nabla\nabla\log\rho\ d\ve x 
              = - \frac 12 E\parenth{\vve G : \nabla\nabla \log\rho},
     \end{eqnarray*}
where integration by parts has been used. So 
	\begin{eqnarray}	\label{eq:dH}
	\dt H = E\parenth{\nabla \cdot \ve F} 
              - \frac 12 E\parenth{\vve G : \nabla\nabla \log\rho}.
	\end{eqnarray}

One of our purposes for this study is to see whether 
the evolution of $H$ can be reconciled to comply with the second law of
thermodynamics, by taking away the effect of phase space volume change, i.e.,
$E\parenth{\nabla \cdot \ve F}$ in this formula. That is to say, 
we would like to see whether $E\parenth{\vve G : \nabla\nabla \log\rho}$
is non-positive. Unfortunately, this need not be true in general.
However, if $\vve G$ is constant in $\ve x$ or, 
in other words, if the noise is additive, then $\vve G$ can be taken
out of the expectation. Integrating by parts,
	\begin{eqnarray*}
	&& E(\vve G : \nabla\nabla\log\rho)
	   = \vve G : \int \rho (\nabla\nabla\log\rho)\ d\ve x \\
	&& \qquad
	   = \vve G : \parenth{-\int \nabla\rho  \nabla\log\rho\ d\ve x}\\
	&& \qquad
	   = - \vve G : \int \rho \nabla\log\rho  \nabla\log\rho\ d\ve x \\
	&& \qquad
	   = - \vve G : E(\nabla\log\rho \nabla\log\rho)	\\
	&& \qquad
	   = - E(\nabla\log\rho \cdot \vve G \cdot \nabla\log\rho).
	\end{eqnarray*}
Because $\vve G = \vve B\ \vve B^T$ is nonnegatively definite, 
$\nabla\log\rho \cdot \vve G \cdot \nabla\log\rho \ge 0$,
hence 
	\begin{eqnarray}	\label{eq:dH_additive}
	\dt H - E(\nabla\cdot\ve F) = \frac 12
	   E(\nabla\log\rho \cdot \vve G \cdot \nabla\log\rho) \ge 0.
	\end{eqnarray}
That is to say, in this case, systems without phase volume
expansion/contraction in the deterministic limit 
(such as the Hamiltonian system), absolute entropy
is in accordance with the second thermodynamic law.

It is interesting to note that the above formula (\ref{eq:dH_additive})
may be linked to Fisher information if the parameters, say $\mu_i$, of 
the distribution are bound to the state variables in a form of translation 
such as that in a Gaussian process. In this case, one can replace the partial 
derivatives with respect to $x_i$ by that with respect to $\mu_i$.
And, accordingly,
	\begin{eqnarray*}
	E\parenth{\nabla\log\rho \nabla\log\rho} = \vve I,
	\end{eqnarray*}
where $\vve I = (I_{ij})$ is the Fisher information matrix.
So
	\begin{eqnarray}
	\dt H = E\parenth{\nabla\cdot\ve F} + \frac 1 2 \vve G : \vve I.
	\end{eqnarray}

Next look at the relative entropy (\ref{eq:D}).
For the reference density $q$, it is also governed by
the Fokker-Planck equation, which reads
	\begin{eqnarray} 	\label{eq:fokkerq}
	\Dt q = - \nabla\cdot(q\ve F) + \frac 12 \nabla\nabla : (q\vve G).
	\end{eqnarray}
Substituting (\ref{eq:fokker}) and (\ref{eq:fokkerq}) into the identity
	\begin{eqnarray*}
	\Dt {(\rho\log q)} = \Dt\rho \log q + \frac \rho q \Dt q
	\end{eqnarray*}
for $\Dt\rho$ and $\Dt q$, and then integrating 
over $\R^n$, we get
	\begin{eqnarray}	\label{eq:fokker_H0}
	&& - \dt\ {\int_\Omega \rho\log q\ d\ve x}
	  = - \int \parenth{\Dt\rho\log q + \frac\rho q \Dt q} d\ve x	\cr
	&&= - \int\log q \bracket{- \nabla\cdot(\rho\ve F) 
			  + \frac12\nabla\nabla : (\rho\vve G)} d\ve x
	    - \int \frac\rho q \bracket{- \nabla \cdot (q\ve F) 
			  + \frac12\nabla\nabla : (q\vve G)} d\ve x 	\cr
	&&= \int \bracket{\log q \nabla\cdot(\rho\ve F) 
			 + \frac\rho q \cdot(q\ve F)} d\ve x 
	    - \frac 12 \int \bracket{\log q \nabla\nabla : (\rho\vve G)
		 + \frac\rho  q \nabla\nabla : (q\vve G)} d\ve x 	\cr
	&&= \int \bracket{\log q \nabla\cdot(\rho\ve F) + 
	    \nabla\log q \cdot \rho\ve F + \rho \nabla\cdot\ve F} d\ve x
	    - \frac 12 \int \bracket{\log q \nabla\nabla : (\rho\vve G)
		 + \frac\rho  q \nabla\nabla : (q\vve G)} d\ve x 	\cr
	&&= E\parenth{\nabla\cdot\ve F} - \frac 1 2 \bracket{
		\int \log q \nabla\nabla : (\rho\vve G) d\ve x +
	        \int \frac \rho q \nabla\nabla : (q\vve G) d\ve x}.
	\end{eqnarray}
Subtracting (\ref{eq:fokker_H}) from above gives the
time evolution of the relative entropy:
	\begin{eqnarray}
	\dt D = \frac 12 \int  \bracket{
		\parenth{\log\frac\rho q} \nabla\nabla : (\rho\vve G)
		+ \nabla\nabla : (\rho\vve G) 
		- \parenth{\frac \rho q} \nabla\nabla : (q\vve G) 
				       } \ d\ve x.
	\end{eqnarray}
Integrating by parts, and using the compact support assumption,
this becomes
	\begin{eqnarray}		\label{eq:dD}
	\dt D 
	&=& - \frac12 \int\bracket{\nabla\log\frac\rho q
		\cdot\nabla\cdot(\rho\vve G) 
	    - \nabla\parenth{\frac\rho q}\cdot\nabla\cdot(q\vve G)} 
		d\ve x				\cr
	&=& - \frac12 \int \bracket{
		\frac q\rho \nabla\parenth{\frac\rho q} \cdot
		(\nabla\rho\cdot\vve G + \rho\nabla\cdot\vve G)
		- \nabla\parenth{\frac\rho q} 
		  \cdot (\nabla q \cdot\vve G + q \nabla\cdot\vve G)
				  } d\ve x	\cr
	&=& \frac 12 \int \frac 1\rho \nabla\parenth{\frac\rho q}
	    \cdot \vve G \cdot (\rho\nabla q - q\nabla\rho)\ d\ve x\cr
	&=& \frac 12 \int \rho \nabla\parenth{\frac\rho q} 
		\cdot \vve G \cdot \nabla\parenth{\frac q \rho}\ d\ve x \cr
	&=& \frac 12 E\bracket{
	  	\nabla\parenth{\frac\rho q} \cdot \vve G \cdot 
	  	\nabla \parenth{\frac q\rho} } 	\cr
	&=& \frac 12 E\bracket{
		\parenth{\frac q\rho \nabla\parenth{\frac\rho q}} 
		\cdot \vve G \cdot
		\parenth{\frac \rho q \nabla\parenth{\frac q \rho}} } \cr
	&=& \frac 12 E\bracket{
		\nabla\parenth{\log\frac\rho q} \cdot \vve G \cdot
		\nabla\parenth{\log\frac q\rho} }	\cr
	&=& - \frac 12 E\bracket{
		\nabla\parenth{\log\frac\rho q} \cdot \vve G \cdot
		\nabla\parenth{\log\frac\rho q}
				    }.
	\end{eqnarray}
Because of the nonnegative definiteness of $\vve G = \vve B\ \vve B^T$,
the right hand side is always smaller than or equal to zero,
in accordance with the thermodynamic entropy. 
(Notice the negative sign in the definition of $D$; that is to say,
increase in $H$ corresponds to decrease in $D$.)


We have studied the evoluationary laws for absolute entropy $H$
and relative entropy $D$ with respect to dynamical systems. 
For easy reference, the derived formulas are wrapped up here. 
If the system of concern is deterministic, i.e., in the form 
of (\ref{eq:gov}), then
    \begin{eqnarray*}
    \dt H &=&  E \parenth{\nabla\cdot\ve F}, 
      \	\quad\qquad\qquad\qquad\qquad\qquad\qquad (\ref{eq:entropy_gov})\\
    \dt D &=&  0. 			       	\qquad\quad
	\qquad\qquad\qquad\qquad\qquad\qquad\qquad (\ref{eq:re_entropy_gov})
    \end{eqnarray*}
If the system has stochasticity included, as that in (\ref{eq:gov2}), then
	\begin{eqnarray*}	
	\dt H &=& E\parenth{\nabla \cdot \ve F} 
              - \frac 12 E\parenth{\vve G : \nabla\nabla \log\ell_\rho}, 
				\qquad\qquad (\ref{eq:dH}) \\
	\dt D &=& - \frac 12 E\bracket{
		\nabla \ell_{\rho/q} \cdot \vve G \cdot
		\nabla \ell_{\rho/q} },
			\quad\qquad\qquad\qquad\qquad (\ref{eq:dD})
	\end{eqnarray*}
where $\ell_\rho = \log\rho$, $\ell_{\rho/q} = \log\rho/q$,
and $\vve G = \vve B\ \vve B^T$.
Among the four formuas, (\ref{eq:entropy_gov}) was known before, 
the rest were obtained in this letter. From them we see that generally 
absolute entropy does not comply with the second law of
thermodynamics, unless the flow of the deterministic vector field is 
nondivergent (as that in a Hamiltonian system) 
and the noisy perturbation is additive. 
The relative entropy, in contrast, proves to be non-increasing all the time,
in accordance with the second law. The dissipative mechanism has a
form remniscient of the Fisher information.

Of particular interest among the above formulas are those for 
deterministic systems. They have important implications from both 
theoretical and applied points of view. For example, drifter releasing
as one of the oldest methods of studying ocean circulation has built up
for oceanic scientists a huge database; but the drifter trajectories
are usually chaotic and are difficult to analyze. Here
(\ref{eq:re_entropy_gov}) and (\ref{eq:entropy_gov}) may come to
help by offering two constraints.
The former tells that the relative entropy is conserved.
For the latter, the sea water is incompressible and hence the oceanic
flow is divergence free. 
So by (\ref{eq:entropy_gov}) the absolute entropy of these trajectories
is also a constant. Equally this applies to the study 
of atmospheric pollutant dispersion. 
Though the air is compressible, but in an isobaric frame
it is not, and hence atmospheric flows are also divergence free.
So in isobaric coordinates the pollutant trajectories must also conserve
their absolute entropy, as well as the relative entropy.

Relative entropy has an interpretation that it measures the distance 
between two functions $\rho$ and $q$ in the function space 
$L^1(\R^n)$ (i.e., integrable functions)\cite{Cove}, 
although it does not meet all the axioms for a metric.
This interpretation makes the relative entropy conservation law,
namely (\ref{eq:re_entropy_gov}), theoretically very interesting.
To see this, examine a nonlinear system that is sensitive 
to initial perturbations. The sensitivity is quantitatively characterized
by the maximal Lyapunov exponent (MLE), which measures the exponential
growth of the separation of two trajectories closely placed in the
beginning\cite{chaos_book}. More specifically, 
if the Lyapunov exponent is $\lambda$,
and the distance between two trajectories is $\delta(t)$, then
	$\frac {\delta(t)} {\delta(0)} \approx e^{\lambda t}.$
Usually a system is considered as chaotic if the MLE is positive;
corespondingly the predictability is quickly lost. 
Now, the relative entropy conservation law tells 
that, if instead of studying the evolution of the 
state variables $\ve x \in \R^n$, we study the evolution of 
their joint density $\rho \in L^1(\R^n)$, the ``trajectories''
in the new ``phase space'' $L^1(\R^n)$ will have equal separations
all the time. That is to say, although the trajectories of
$\ve x$ may be chaotic, the ``trajectories'' of $\rho(\ve x)$
are not, and the corresponding Lyapunov exponent $\lambda$ will
always be zero. 

The above observation is expected to have important implications
in the the active research field, ensemble prediction. 
Realizing the limited predictability of nonlinear
dynamical systems, during the past decades there has been a surge
of interest in ensemble prediction, for instance, ensemble weather
prediction\cite{ensemble}. 
The implication is two-fold. Firstly, the law rationalizes the 
prediction technique, in that it assures the insensitivity of 
the distribution to initial conditions.
In this sense, the conservation law may be taken
as the theoretical basis of ensemble prediction.
Seondly, the law imposes a constaint on the 
numerical schemes designed for prediction.
We know, in approximating the differential operators in a 
(deterministic) system for numerical computation, 
the underlying physics is, more or less, changed.
For instance, artifical damping at each step
may be used to ensure numerical stability;
stochasticity may be deliberately introduced to parameterize the processes
that cannot be resolved by the model grids;
the ensemble size may be too small to cover the sample space, and so forth.
For high dimensional problems such as weather forecast, the latter
is particularly severe, as the integration is very expensive.
All these may lead to a non-conservative relative entropy, and hence 
the resulting prediction may not be able to reflect the real 
statistical physics underyling the system. How to design a 
relative entropy conservative scheme is, therefore, of interest for 
ensemble predictions with high dimensional systems. We leave
this to future studies.

\begin{acknowledgments}
Discussions with Richard Kleeman are appreciated.
\end{acknowledgments}

\bibliography{info}

\end{document}